\documentclass[11pt,a4paper]{article} 
\pdfoutput=1
\usepackage{jcappub}

\title{Observational consequences of chaotic inflation with nonminimal coupling to gravity}
\author[a]{Andrei Linde,} \author[a]{Mahdiyar Noorbala} \author[b]{and Alexander Westphal}

\affiliation[a]{Stanford Institute for Theoretical Physics and Department of Physics, \\ Stanford University, Stanford, CA 94305 USA} \affiliation[b]{Deutsches Elektronen-Synchrotron DESY, Theory Group, D-22603 Hamburg, Germany} 
\emailAdd{alinde@stanford.edu} \emailAdd{noorbala@stanford.edu} \emailAdd{alexander.westphal@desy.de}
\abstract{Recently there was an extensive discussion of Higgs inflation in the theory with the potential ${\lambda\over 4}(\phi^{2}-v^{2})^{2}$ and nonminimal coupling to gravity ${\xi\over 2}\phi^{2}R$, for $\xi \gg 1$ and $v\ll 1$. We extend this investigation to the theories ${m^{2}\over 2}\phi^{2}$ and ${\lambda\over 4}(\phi^{2}-v^{2})^{2}$ with arbitrary values of $\xi$ and $v$ and describe implementation of these models in supergravity.  We analyze observational consequences of these models and find a surprising coincidence of the inflationary predictions of the model ${\lambda\over 4}(\phi^{2}-v^{2})^{2}$ with $\xi <0$ in the limit $|\xi|v^{2}\to 1$ with the predictions of the Higgs inflation scenario for $\xi \gg 1$.
}
 \arxivnumber{1101.2652}
\def\be{\begin{equation}}
\def\ee{\end{equation}}
\def\ba{\begin{eqnarray}}
\def\ea{\end{eqnarray}}
\def\K{K{\"a}hler}
\newcommand{\eq}[1]{Eq.~(\ref{#1})}
\newcommand{\fig}[1]{Fig.~\ref{#1}}
\usepackage{amssymb,graphicx,subfigure}

\begin{document}
\begin{flushright}
DESY 11-013
\end{flushright}
%\vfil
\maketitle

\section{Introduction}
The chaotic inflation scenario  \cite{Linde:1983gd} is one of the simplest and most general versions of inflationary cosmology. It can be realized in many theories with sufficiently flat potentials, including the models with $V \sim \phi^{n}$.  However, for a long time it seemed very difficult to implement chaotic inflation in superstring theory and in supergravity. This situation changed when a simple realization of the chaotic inflation scenario in supergravity was proposed in  \cite{Kawasaki:2000yn}, and a class of chaotic inflation models was developed in the context of string theory \cite{Silverstein:2008sg}. In this class of models, one could have an inflaton potential $\sim \phi^{2/3}$, or $\phi^{{4/5}}$, or $\phi$, predicting a much smaller value of $r$.

Recently a broad class of models of chaotic inflation in supergravity was proposed in  \cite{Kallosh:2010ug}. In these models one can obtain chaotic inflation with an arbitrary inflaton potential $V(\phi)$. In these models one can also implement the curvaton scenario \cite{Linde:1996gt} with a controllable degree of non-gaussianity  \cite{Demozzi:2010aj}.

Moreover, this new class of supergravity models can describe an inflaton field which is nonminimally coupled to gravity, with coupling ${\xi\over 2} \phi^{2}R$ \cite{Einhorn:2009bh,Kallosh:2010ug}. Recently there was a wave of interest to such models after the suggestion to implement chaotic inflation in the standard model of electroweak interactions, with the Higgs field nonminimally coupled to gravity playing the role of the inflaton field \cite{Salopek:1988qh,Bezrukov:2007ep}, see also \cite{Spokoiny:1984bd}. However, the traditional textbook formulation of supergravity did not provide any description of scalar fields  nonminimally coupled to gravity. This could suggest that the models with nonminimal coupling to gravity are outdated, because they are incompatible with modern developments in particle physics based on supersymmetry. Fortunately, this problem was solved in \cite{Einhorn:2009bh,Kallosh:2010ug}, where a consistent generalization of supergravity describing scalar fields nonminimally coupled to gravity was developed and a supersymmetric generalization of the inflationary model of Refs. \cite{Salopek:1988qh,Bezrukov:2007ep} was proposed. 

These developments prompted us to return to an investigation of various versions of chaotic inflation with the inflaton field nonminimally coupled to gravity. In this paper we will consider two basic models:  the simplest model with a quadratic potential ${m^{2}\over 2}\phi^{2}$, and  the model with the standard spontaneous symmetry breaking potential ${\lambda\over 4}(\phi^{2}-v^{2})^{2}$. Our goal is to analyze these models at the classical level, describe the way to incorporate them  in supergravity, and study the stability of the inflationary trajectory in these theories. We will also calculate the spectral index $n_{s}$ and the tensor-to-scalar ratio $r$ for various values of the parameter $\xi$ describing the nonminimal coupling of the inflaton field to gravity in these models. We will show that by changing parameters of these models one can cover a significant part of the space of parameters ($n_{s},r$) allowed by recent observational data. 
Finally, we will show that the inflationary predictions of the model ${\lambda\over 4}(\phi^{2}-v^{2})^{2}$ with $\xi <0$ in the limit $|\xi|v^{2}\to 1$ coincide with the predictions of the Higgs inflation scenario for $\xi \gg 1$ and $v\ll 1$.

\section{Canonical Field in Einstein Frame}
The  Lagrangian of the inflaton field in the models discussed in the Introduction describes a scalar field $\phi$ with a canonical kinetic term but with a {\em non-minimal} coupling to gravity:
\begin{equation}\label{LJ}
L = \frac{1}{2} \sqrt{-g_J} \left[ (1+\xi \phi^{2}) R_J - g_J^{\mu\nu} \partial_\mu \phi \partial_\nu \phi - 2 V_J(\phi) \right].
\end{equation}
The index $J$ is to emphasize that this is a Jordan frame: the gravity part of the Lagrangian is different from the Einstein Lagrangian $\frac{1}{2}  \sqrt{-g} R[g]$ and has a local function $1+\xi \phi^{2}$  in front. It is possible to proceed with this form of the Lagrangian but we will be able to use the familiar equations of general relativity, the inflationary solutions and the standard slow-roll analysis if we switch to the Einstein frame and use variables  that show {\em minimal} coupling between the scalar field and gravity.

To this end, we change variables from $g_J$ and $\phi$ to a conformally related metric $g_E^{\mu\nu} = (1+\xi \phi^{2})^{{-1}}  g_{J}^{\mu\nu}$ and a new  field $\varphi$ related to the field $\phi$ as follows:
\begin{equation}\label{trans}
\left( \frac{d\varphi}{d\phi} \right)^2 = \frac{1+\xi \phi^2 + 6\xi^{2} \phi^2}{(1+\xi \phi^2)^2} =: \frac{1}{Z(\phi)}. 
\end{equation}
 In new variables, the Lagrangian, up to a total derivative, is given by
\begin{equation}\label{LE}
L = \frac{1}{2} \sqrt{-g_E} \left[ R_E - g_E^{\mu\nu} \partial_\mu \varphi \partial_\nu \varphi - 2 V_E(\phi(\varphi)) \right].
\end{equation}
where the potential in the Einstein frame is 
\begin{equation}\label{eframe}
V_E(\phi) = \frac{V_J(\phi)}{(1+\xi \phi^2)^2},
\end{equation}
The canonically normalized Einstein-Hilbert term of the action shows that $g_E$ is the metric in the Einstein frame.  Therefore we can use all results of the standard slow-roll analysis.

In doing the slow-roll analysis we should note that it is $\varphi$ and not $\phi$ that has a canonical kinetic term.  Therefore the slow-roll parameters are
\begin{equation}
\epsilon = \frac{1}{2} \left( \frac{V_\varphi}{V} \right)^2, \qquad \eta = \frac{V_{\varphi\varphi}}{V}.
\end{equation}
(From now on, whenever we use $V$ without a subscript or the word potential we mean the Einstein frame potential $V_E$.)   The same quantities can be formally defined for the field $\phi$ but are not of physical meaning.  However, they are easier to compute and are related to the physical slow-roll parameters via
\begin{equation}\label{phys-eps-eta}
\epsilon = Z \epsilon_\phi, \qquad \eta = Z \eta_\phi + \operatorname{sgn}(V') Z' \sqrt{\frac{\epsilon_\phi}{2}},
\end{equation}
where prime means $d/d\phi$. 
Similarly, we can find the number of $e$-folds by
\begin{equation}
N \approx \left| \int \frac{d\varphi}{\sqrt{2\epsilon}} \right| = \left| \int \frac{d\phi}{Z(\phi) \sqrt{2\epsilon_\phi}} \right|.
\end{equation}
This enables us to do almost all computations in terms of $\phi$ without having to solve \eq{trans} to find $\varphi$.  But we should not forget the fact that $\phi$ has a non-canonical kinetic term $-\frac{1}{2} Z^{-1} g_E^{\mu\nu} \partial_\mu \phi \partial_\nu \phi$.  So when we qualitatively analyze the motion of the field, in analogy with a particle with Lagrangian $\frac{1}{2}m{\dot{x}}^2-V(x)$, we need to associate a variable `rolling mass' $Z^{-1}(\phi)$ to the field $\phi$.\footnote{This `rolling mass' should not be confused with the usual mass $V''$ of the field.}  This doesn't change the direction of rolling but modifies the velocity of the field such that with heavier rolling mass (smaller $Z$) the motion becomes slower (smaller $\epsilon$ and $\eta$) as indicated by \eq{phys-eps-eta}.

In the small field limit, $\phi = \varphi$, but at large values of the fields their relation to each other is more complicated. In particular, for $\xi > 0$ and $\phi \gg {1/\sqrt\xi}$ (or, equivalently, $\varphi \gg 1$), one has 
\be\label{positive}
\phi = {1\over \sqrt \xi} e^{{\varphi\over\sqrt 6}} \ .
\ee
Therefore if the potential $V(\phi)$ is a polynomial of $\phi$, at $\phi \gg {1/\sqrt\xi}$ (i.e. at $\varphi \gg 1$) it depends on the canonically normalized field exponentially.

For negative $\xi$, the potential becomes singular when $\phi$ approaches $1/\sqrt{|\xi|}$, beyond which the effective gravitational coupling constant $G_N ={1\over 8 \pi(1+\xi\phi^{2})}$ would change sign (antigravity regime).  However, in terms of the canonically normalized field $\varphi$, the singularity is never reached, because it corresponds to infinitely large values of $\varphi$. That is why in this scenario one should not worry about the possibility to cross the boundary between gravity and antigravity, or even to approach this boundary \cite{Linde1979,Starobinsky1981}.

One can especially clearly see it if one considers a vicinity of the would-be singularity, where $1-|\xi| \phi^{2} \ll 1$. In this regime 
\be\label{negative}
\phi = {1\over \sqrt {|\xi|}}\left(1-  {1\over 2}\,e^{-\sqrt{2\over 3}\,\varphi}\right) \ .
\ee
Therefore $1-\sqrt{|\xi| }\phi = {1\over 2}\, e^{-\sqrt{2\over 3}\,\varphi}$, which means that the limit $\phi \to 1/\sqrt{|\xi|}$ corresponds to the limit $\varphi \to \infty$.
These considerations will help us to understand the properties of the inflaton potential in the Einstein frame.

Before investigating  cosmological implications of this class of models, we would like to make a short detour and discuss implementation of these models in the context of supergravity. The results obtained in the paper do not depend on the possibility to implement the inflationary models with such potentials in supergravity, but supergravity allows to look at such models and evaluate them from an entirely different perspective. The readers interested only in observational consequences of the models with nonminimal coupling to gravity may skip the next section and go straight to section \ref{quadratic}. 

\section{Supergravity perspective}\label{SUGRA}

The new class of chaotic inflation models in supergravity \cite{Kallosh:2010ug} describes two fields, $S$ and $\Phi$, with the superpotential
\be 
W = S\, f(\Phi), 
\ee
where $f(\Phi)$ is a real holomorphic function such that $\bar f(\Phi) = f(\Phi)$. 
The  \K\, potential in the simplest version of this model is given by
\be
\mathcal{K} = -3\log \Bigl[ 1 - \frac{1}{3}(\Phi \bar\Phi + S \bar S)+ {\chi\over 4 }(\Phi^2+ \bar\Phi^2) +\zeta (S\bar S)^2/3 \Bigr] .
\label{Ka}
\ee
The term $\zeta (S\bar S)^2/3$ is added for stabilization of the inflationary trajectory, to ensure that the fields  $S$ and ${\rm Im}\, \Phi$  vanish during inflation: $S= {\rm Im}\, \Phi =0$. The field $\phi = \sqrt 2 {\rm Re\, \Phi}$ plays the role of the inflaton field in this scenario. According to  \cite{Einhorn:2009bh,Kallosh:2010ug}, the evolution of this field is described by the theory of the scalar field $\phi$ nonminimally interacting with gravity, with $\xi=-\frac{1}{6}+\frac{\chi}{4}$. The Jordan frame potential of this field is 
\be
V_J =f^{2}({\rm Re}\, \Phi) \ .
\ee
It is convenient to present the result in terms of the real  fields $s$, $\alpha$, $\phi$ and $\beta$, related to the complex fields through
\be
S  ={1\over\sqrt 2}(s+i\alpha)\, , \qquad \Phi  ={1\over\sqrt 2}(\phi+i\beta) \ \ .
\label{cart}\ee
The potential of the inflaton field $\phi$ is then
\be
V_J(\phi) =f^{2}(\phi/\sqrt 2)\  ,
\ee
and the Einstein frame potential is given by Eq. (\ref{eframe}).

Thus, in this context one has a functional freedom of choice of the inflaton potential, determined by the function $f(\Phi)$. However, this scenario works only if one can actually ignore the cosmological evolution of the fields $S$ and ${\rm Im}\, \Phi$, by fixing them at their values $S= {\rm Im}\, \Phi =0$ during inflation. Therefore we must check whether the trajectory $S= {\rm Im}\, \Phi =0$ corresponds to the minimum of the potential in the directions $S$ and ${\rm Im}\, \Phi$. In our scenario, the potential has the same curvature in the directions $s$ and $\alpha$, so one should check whether $m^{2}_{\beta},m^{2}_{s} >0$. We will impose a stronger condition, $m^{2}_{\beta},m^{2}_{s} \gtrsim H^{2}$, to avoid production of inflationary perturbations of any fields except for the inflaton field $\phi$.

These conditions can be analyzed along the lines of \cite{Kallosh:2010ug}, where it was shown that for the theories with $\xi = 0$ and arbitrary  function $f(\Phi)$ one can always stabilize the inflationary trajectory $S= {\rm Im}\, \Phi =0$ by a proper choice of the parameter $\zeta$.

An investigation of this issue for $\xi \not = 0$ is similar, but more complicated, because  $m^{2}_{\beta}$ and $m^{2}_{s}$ depend on $\xi$:
\begin{equation}
m^2_{\beta} =  { {4\over 3}(1+3\xi)f^{2} + (1+\xi\phi^2)(({\partial_{\Phi}f})^{2}  -f\,  \partial^{2}_{\Phi}f)\over (1+\xi\phi^2)(1+\xi\phi^2 +6\xi^2\phi^2)},
\end{equation}
\begin{equation}
m^2_{s} =  {A(\phi)  \over (1+\xi\phi^2)^2(1+\xi\phi^2 +6\xi^2\phi^2)}. 
\end{equation}
where
\ba
A(\phi) &=& {2\over 3}(-1-\xi\phi^2+6\xi^2\phi^2 +6\zeta(1+2\xi\phi^2 +6 \xi^3 \phi^4\nonumber\\ &+&\xi^2\phi^2(6+\phi^2)))f^{2} -4\sqrt 2 \xi\phi(1+\xi\phi^2)f\,  \partial_{\Phi}f\nonumber\\&+& (1+\xi\phi^2)^2 (\partial_{\Phi}f)^{2}
\ea

In this paper we studied two basic models. The first one corresponds to   $f(\Phi) = m\Phi$, which leads to the quadratic inflaton potential ${m^{2}\over 2}\phi^{2}$. The second choice is $f(\Phi) = \sqrt{\lambda} (\Phi^2 - v^2/2)$, which leads to the inflaton potential ${\lambda\over 4}(\phi^{2}-v^{2})^{2}$. 
Investigation of stability in the supergravity versions of these models is straightforward but quite involved. Here we will skip the details and briefly summarize our main results.

First of all, we verified that the inflationary trajectory is stable with respect to generation of the field $\beta \sim {\rm Im\, \Phi}$. However, the situation with the field $S$ is more complicated. Just as in the models studied in \cite{Einhorn:2009bh}, stabilization requires the existence of the stabilizing term 
$\zeta (S\bar S)^2/3$ in the \K\, potential. For $\xi >0$, stabilization can be achieved by taking $0<\zeta \lesssim O(1)$. Similar conclusion is valid for $\xi<0$ as well, with one caveat. 

As we  discussed in the previous section, the point $|\xi| \phi^{2} = 1$ for $\xi <0$ corresponds to a singularity, where the effective gravitational constant becomes infinite and changes its sign. When $|\xi| \phi^{2}$ increases towards $1$, which corresponds to the vicinity of this dangerous point, or to the infinitely large values of the canonically normalized inflaton field $\varphi$, stabilization of the field $S$ requires $\zeta \gg 1$. This is not particularly attractive, but appearance of large parameters in the \K\, potential is not a novelty in this scenario. Indeed,  the Higgs inflation scenario with $ \xi\gg 1$ \cite{Bezrukov:2007ep} also requires introduction of a very large coefficient $\chi$ in the \K\, potential (\ref{Ka}) \cite{Einhorn:2009bh}.

\section{Basic models and their extensions}
In this paper we will concentrate on observational consequences of two basic models, with potentials $\frac{1}{2}m^2\phi^2$ and ${\lambda\over 4}(\phi^{2}-v^{2})^{2}$, and extend their analysis for the case with $\xi \not = 0$. To put our results in a proper perspective, we will briefly remember what is known about observational consequences of some closely related models.

  \begin{figure}[t!]
\centering
\includegraphics[scale=0.4]{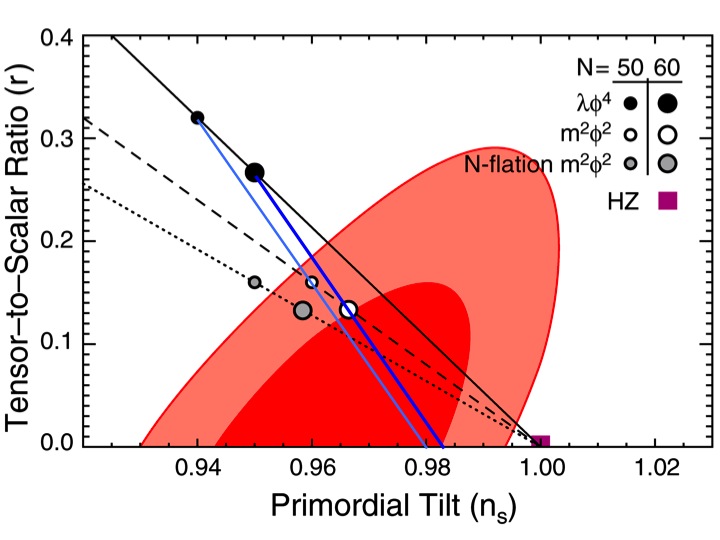}
\caption{Predictions of the chaotic inflation model with the potential $V\sim \phi^{2\alpha}$ for different $\alpha$ in the range of $0< \alpha <2 $ are shown by two blue lines corresponding to $N = 60$ and $N = 50$. Our results are shown on top of the WMAP results.}
\label{WMAP2}
\end{figure}

1) $V \sim \phi^{2\alpha}$, $\xi = 0$. For this model one has $1-n_{s} =(1+\alpha) N^{-1}$, $r = 8\alpha/N$, where $N$ is the number of e-folds.
 In this case $1-n_{s} = 1/N + r/8$  \cite{Silverstein:2008sg}. The results are shown by two blue straight lines in Fig.~\ref{WMAP2}. Not surprisingly, these lines go through the circles corresponding to the theories with potentials proportional to $\phi^{2}$ and $\phi^{4}$; these lines continue going down when $\alpha$ decreases.

\begin{figure}[t]
\centering
\includegraphics[scale=0.2]{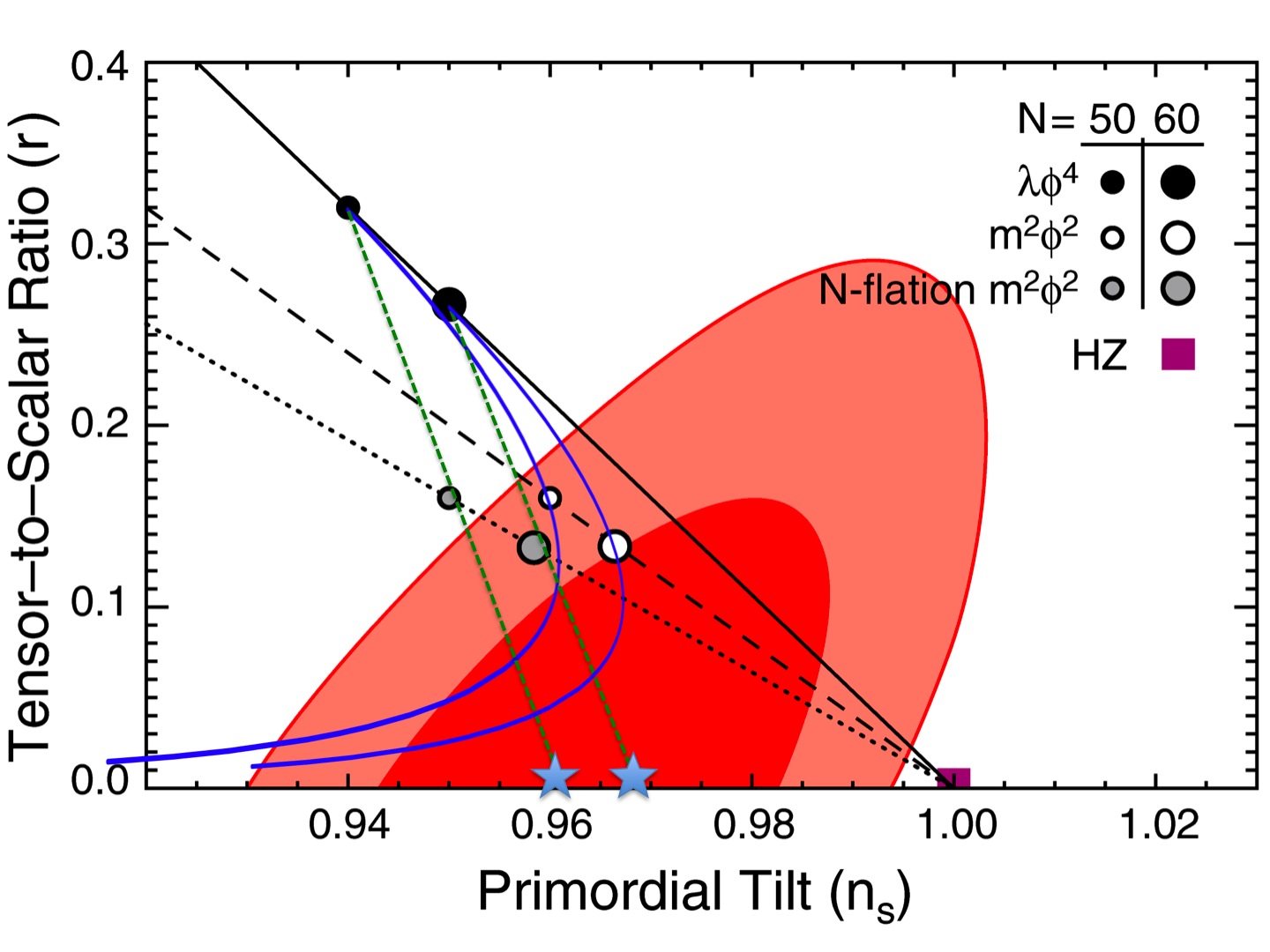}
\caption{Predictions of the model with the potential ${\lambda\over 4} (\phi^2-v^2)^2  $ are bounded by the two blues lines corresponding to the number of e-foldings $N = 50$ and $N = 60$ \cite{Kallosh:2007wm}. The blue stars correspond to this model with $v = 0$ and the inflaton nonminimally coupled to gravity with $\xi \gg 1$ for $N = 50$ and $N = 60$ \cite{Bezrukov:2007ep,Einhorn:2009bh}. The green dashed lines describe predictions of this model for $v = 0$ and for various values of $\xi >0$ \cite{Okada:2010jf}, for $N = 50$ and $N = 60$.}
\label{WMAP}
\end{figure}

2) $V \sim (\phi^{2}-v^{2})^{2}$, $\xi = 0$. The results depend on the value of $v$, and also on the initial conditions for the field $\phi$. If inflation occurs at $\phi > v$, and $v$ is not too large, the results coincide with those in the model with the potential $\phi^{4}$, see Fig.~\ref{WMAP}.  For $v > O(10)$, the last stages of inflation occur not far from the minimum of the potential, where it looks quadratic. Therefore in this regime $n_{s}$ and $r$ coincide with their value for the quadratic potential. This property remains valid for $v > O(10)$ even if the field falls from $\phi < v$. However, inflation is also possible for  $\phi<v$ and $1\lesssim v \lesssim O(10)$, in which case $n_{s}$ and $r$ can be much  smaller, as shown on the lower branch of the blue line in Fig.~\ref{WMAP}, below the white circle.

The green lines in Fig.~\ref{WMAP} show $n_{s}$ and $r$ for the theory with the potential $\phi^{4}$ with $\xi > 0$. This potential coincides with our potential $\sim (\phi^{2}-v^{2})^{2}$ in the limit $v = 0$ \cite{Okada:2010jf}.  

The investigation which we are going to perform should generalize the previously obtained results for the theory $m^{2}\phi^{2}/2$ with nonminimal coupling $\xi$, and  for the theory  ${\lambda\over 4} (\phi^2-v^2)^2  $ with arbitrary values of $v$ and $\xi$.

\section{Quadratic Potential}\label{quadratic}
\begin{figure}[ht]
\centering
\includegraphics[scale=0.7]{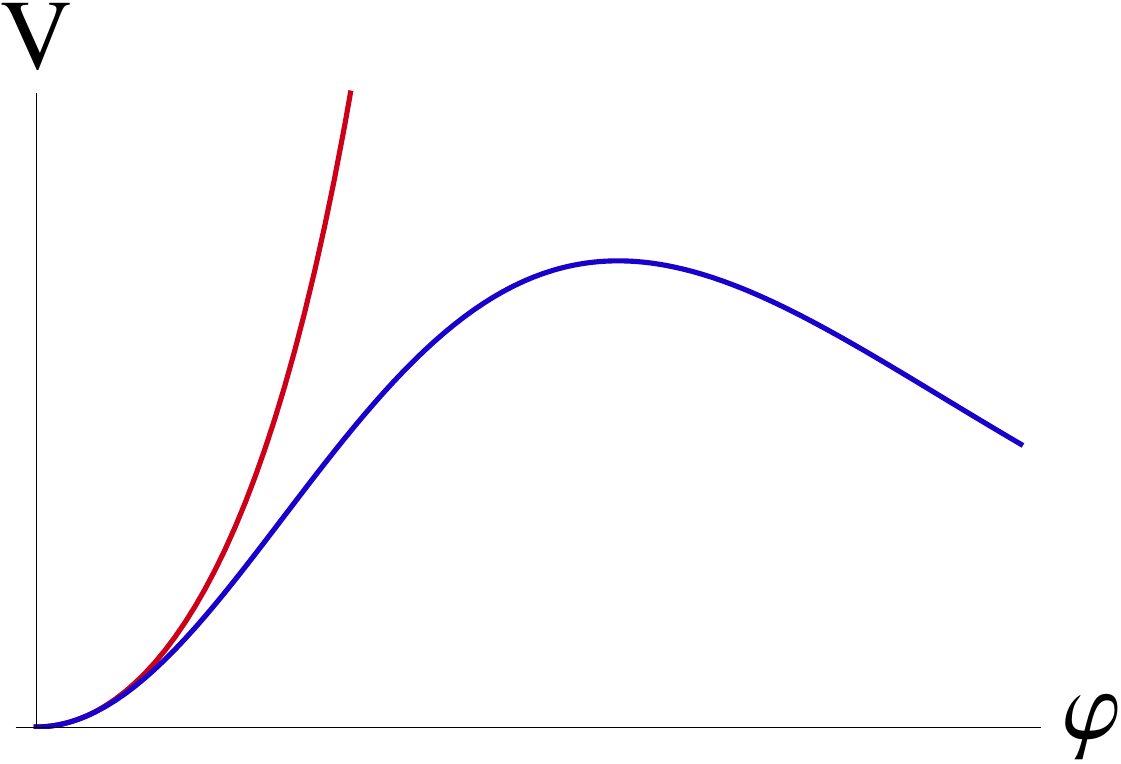}
\caption{The quadratic potential $V$ as a function of the canonically normalized inflaton field $\varphi$ in the Einstein frame. The right, blue curve corresponds to $\xi = 1$, the left, red curve corresponds to $\xi = -1$. Note that the potential for the negative $\xi$ is very steep, but it does not contain any singularity, which appears when one expresses this potential in terms of the original scalar field $\phi$ as in Eq. (\ref{V2}). }
\label{fig:V2}
\end{figure}

We begin with the investigation of  the quadratic potential $V_J=\frac{1}{2}m^2\phi^2$ in the Jordan frame, which leads to the Einstein frame potential
\begin{equation}\label{V2}
V_E = \frac{m^2 \phi^2/2}{(1+\xi\phi^2)^2} \ .
\end{equation}
This equation immediately reveals an unusual behavior of the potential. At small $\phi$, the potential behaves as the original quadratic potential. However, for $\xi > 0$   it has a maximum at $\xi \phi^{2} = 1$, see the blue line in \fig{fig:V2}. If the field begins its evolution beyond this maximum, it continues rolling forever, as in the theory of quintessence. The normal inflationary regime which ends by oscillations near $\phi = 0$ occurs only if the evolution occurs to the left of the maximum, i.e., at $\xi \phi^{2} < 1$. However, inflation may begin close to the maximum of this potential  Ref.~\cite{Sakai:1998rg}.

Meanwhile for $\xi < 0$, there is a singularity of the potential at  $\xi \phi^{2} = -1$. Once again, one can ignore the regime that begins at large $\phi$, beyond the singularity, because it describes the universe in the antigravity regime. One can especially clearly see it if one plots $V_{E}$ not as a function of $\phi$, but as a function of the canonically normalized field $\varphi$, see the red line in Fig. \ref{fig:V2}.

The physical slow-roll parameters are
\begin{equation}
\epsilon = \frac{2{\left( 1 - \xi{\phi}^{2}\right) }^{2}}{{\phi}^{2}\left( 1 + \xi{\phi}^{2} + 6{\xi}^{2}{\phi}^{2}\right) },
\end{equation}
\begin{equation}
\eta = \frac{2\left( 12{\xi}^{4}{\phi}^{6}+2{\xi}^{3}{\phi}^{6}-36{\xi}^{3}{\phi}^{4}-5{\xi}^{2}{\phi}^{4}-6\xi{\phi}^{2}+1\right) }{{\phi}^{2}{\left( 1 + \xi{\phi}^{2} + 6{\xi}^{2}{\phi}^{2}\right) }^{2}}.
\end{equation}
The number of $e$-folds that it takes for the field to roll from $\phi_*$ (when the primordial fluctuations were formed) to $\phi_e$ (when inflation ends) is
\begin{equation}\label{N2}
N \approx \frac{3}{4} \log \frac{1 - \xi^2 \phi_e^4}{1- \xi^2 \phi_*^4} + \frac{1}{4\xi} \log \frac{1-\xi\phi_e^2}{1-\xi\phi_*^2}.
\end{equation}

\begin{figure}[ht]
\centering
\includegraphics[scale=.41]{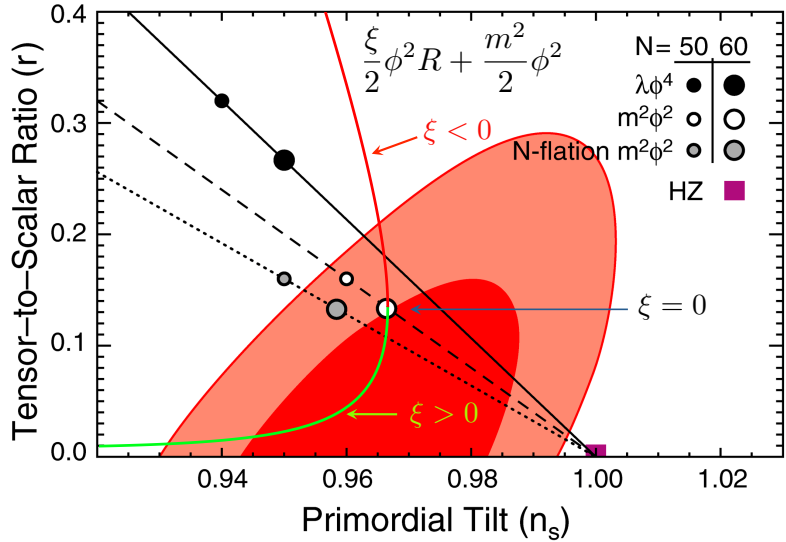}
\caption{$(n_s,r)$ at $N=60$ from the quadratic potential \eq{V2}.  The $\xi=0$ point, corresponding to the simple $m^2\phi^2/2$ potential with minimal gravitational coupling, separates the $\xi<0$ (red, top) and $\xi>0$ (green, bottom) cases.}
\label{fig:nr2}
\end{figure}

Although we have analytic results for slow-roll parameters, it is not easy to solve them to obtain $\phi_*$ at which the observables $n_s$ and $r$ should be evaluated.  Instead we proceed numerically by finding $\phi_e$ (the point where either $\epsilon<1$ or $\eta<1$ fails) and then going $N=60$ $e$-folds back to obtain $\phi_*$ while making sure that the slow-roll parameters remain small in this range of $\phi$.  The result is that inflationary trajectories via slow roll approximation are feasible only for  $-10^{-2} \lesssim \xi \lesssim 10^{-1}$.  Note that by varying $\xi$ from zero we deviate from the point $(n_s,r) = (1-\frac{2}{N}, \frac{8}{N})$ (corresponding to the $m^2 \phi^2 / 2$ potential) and obtain a one-dimensional curve (corresponding to the one parameter $\xi$) in the $n_s$-$r$ plane; see \fig{fig:nr2}.  For large and negative $\xi$ the potential in Fig.~\ref{fig:V2} (the red line) becomes like an exponential barrier within a distance $\Delta \varphi \sim 1$ from the origin.  Hence $N \gtrsim 60$ is accommodated only for $|\xi| \ll 10^{-2}$.  Even when $N \gtrsim 60$ is possible the predicted $n_s$ and $r$ may be well outside the current bounds.  In fact, for $N=60$ the part of the curve in \fig{fig:nr2} that lies within the WMAP $2\sigma$ data corresponds to an even smaller range $-7\times10^{-4} \lessapprox \xi \lessapprox 7\times10^{-3}$.   
In order to obtain  the observed amplitude of density perturbations $\delta_\rho \approx 5\times10^{-5}$ one should increase the mass from $m=3\times10^{-6}$ at $\xi=7\times10^{-3}$ to $m=6\times10^{-6}$ at $\xi=0$, followed by an insignificant decrease in $m$ to reach $\xi=-7\times10^{-4}$ (in fact, only at $\xi=-2.6\times10^{-3}$ which is far outside the observed region, we do need $m=3\times10^{-6}$ again).  In either case, the mass remains within the same order of magnitude of the minimal coupling case.
We see that negative $\xi$ is almost ruled out at $1\sigma$ but a range of positive values of $\xi$ is available to account for forthcoming data down to $r\sim5\times10^{-3}$, albeit with a fine-tuning of at least ${\cal O}(10^{-2})$ in $\xi$.

\section{Quartic Potential}
Let us now consider the standard potential with spontaneous symmetry breaking, which looks -- with non-minimal coupling -- as follows \begin{equation} \label{V4}
V_E = \frac{\lambda}{4} \frac{(\phi^2 - v^2)^2}{(1+\xi\phi^2)^2},
\end{equation}
in the Einstein frame. 

\begin{figure}[h!]
\centering
\subfigure[]{
\includegraphics[scale=.62]{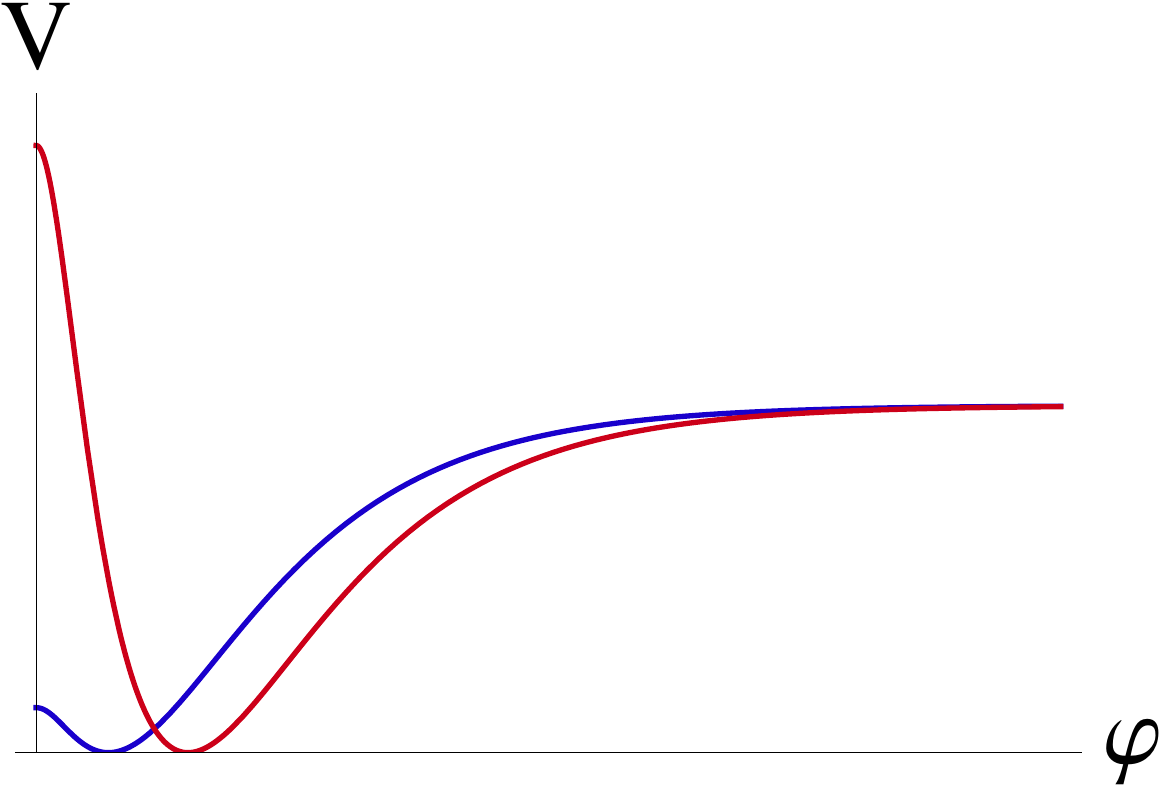}
\label{fig:V4}}
\subfigure[]{
\includegraphics[scale=.65]{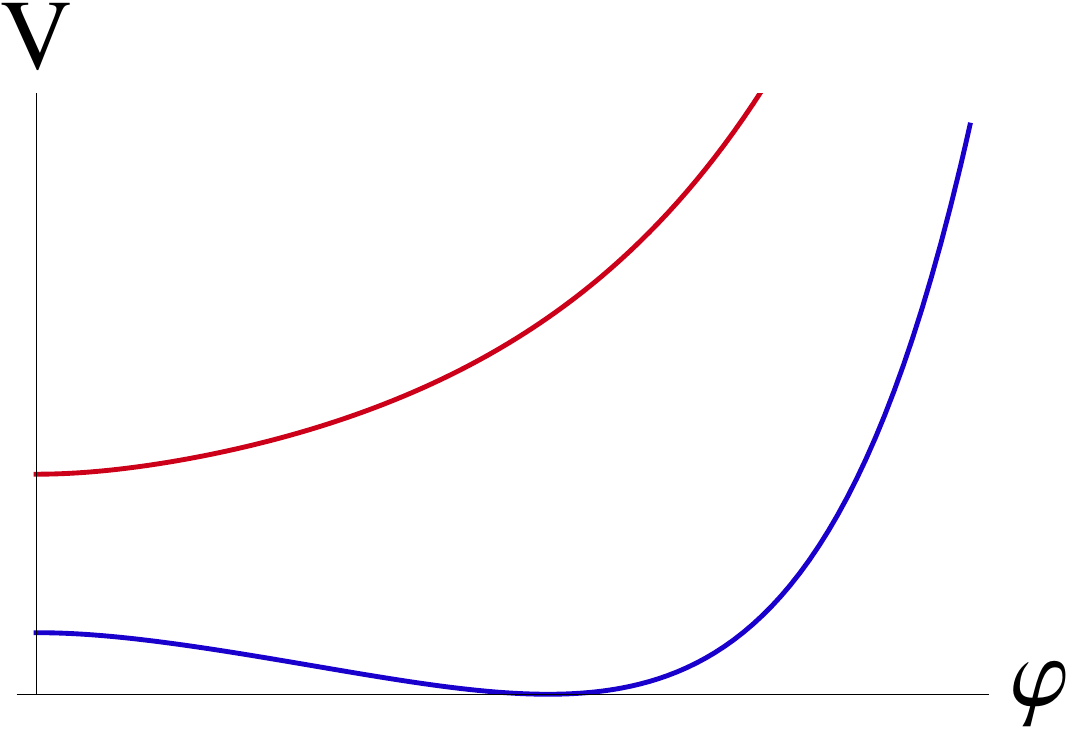}
\label{fig:V4n}}
\caption{Einstein frame potential (\ref{V4}) as a function of the canonically normalized field $\varphi$.  The blue lines correspond to $v < 1$, and the red lines correspond to $v > 1$.  (a) $\xi=1$. (b) $\xi=-1$.  Notice that the potential with $\xi = -1$, $v >1$ has a minimum at a very large value of the potential (the minimum at $V =0$ appears in the antigravity regime beyond the singularity), so this potential is unsuitable for the description of inflation in our universe.}
\end{figure}

To have a better intuitive understanding of the properties of the potential in the Einstein frame, we plot it as a function of the canonically normalized field $\varphi$ in Figs.~\ref{fig:V4} and \ref{fig:V4n} (with $\xi = 1$ and $-1$, respectively) for two different values of $v$ ($v < 1$, blue line;  and $v > 1$, red line).

The slow-roll parameters are now given by
\begin{equation}
\epsilon = \frac{8{\left( 1 + \xi{v}^{2}\right) }^{2}{\phi}^{2}}{{\left( \phi^2-v^2\right) }^{2}\left( 1 + \xi{\phi}^{2} + 6{\xi}^{2}{\phi}^{2}\right) },
\end{equation}
\be
\eta = -\frac{4A\left( 1 + \xi{v}^{2}\right)}{{\left( \phi^2-v^2\right) }^{2}\left( 1 + \xi{\phi}^{2} + 6{\xi}^{2}{\phi}^{2} \right)^{2}} ,
\ee
where
\be
A =  12{\xi}^{3}{\phi}^{6}+2{\xi}^{2}{\phi}^{6}-24{\xi}^{3}{v}^{2}{\phi}^{4}-4{\xi}^{2}{v}^{2}{\phi}^{4}-12{\xi}^{2}{\phi}^{4}-\xi{\phi}^{4}-3\xi{v}^{2}{\phi}^{2}-3{\phi}^{2}+{v}^{2}\ .
\ee
The number of $e$-folds is given by
\be
N \approx \left| \frac{3}{4} \log \frac{1+\xi\phi_e^2}{1+\xi\phi_*^2} + \frac{1}{4\left( 1+\xi{v}^{2}\right) } \left[\frac{1}{2} \left( 1 + 6\xi \right)  (\phi_*^{2} - \phi_e^2) + {v}^{2}\log\frac{\phi_e}{\phi_*}) \right] \right|.
\ee

\begin{figure}[ht]
\centering
\includegraphics[scale=.45]{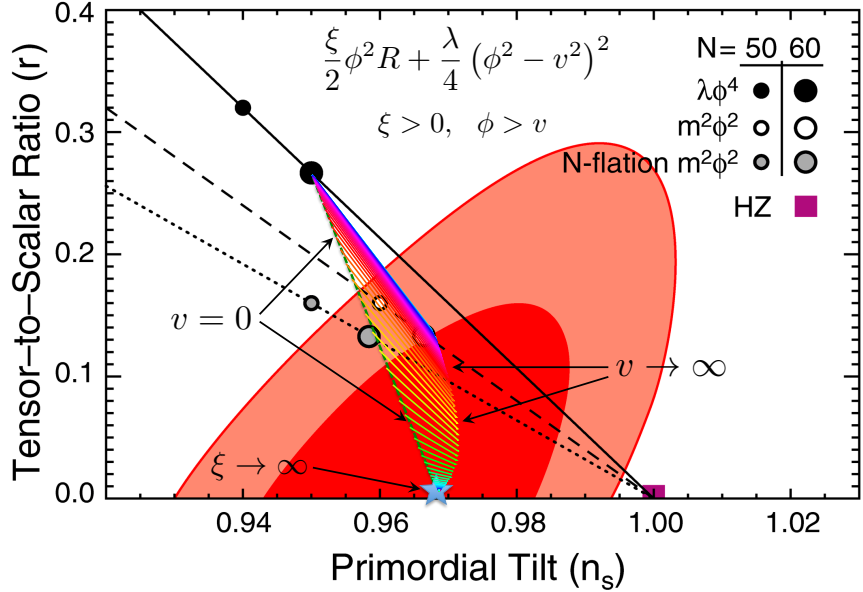}
\caption{$(n_s,r)$ at $N=60$ for the quartic potential \eq{V4}  with positive $\xi$ in the case with inflation at $\phi > v$. Each red and yellow colored line corresponds to a fixed value of $\xi $. For all $\xi$, the colored lines begin at the green dashed line $v =0$, which was found in \cite{Okada:2010jf}. Along each colored line $v$ grows from 0 on left to $\infty$ on right.  For $\xi\gtrapprox0.0027$ each line segment lies entirely within the $1\sigma$ region, regardless of the value of $v$. The limiting point of this set of curves, for $\xi \to \infty$ and any $v$, coincides with the result of the Higgs inflation model with $\xi \gg 1$ and $v \ll 1$ studied in  \cite{Bezrukov:2007ep}. This point is shown by a gray star.  }
\label{fig:nr4-pos-lf}
\end{figure}

\begin{figure}[ht]
\centering
\includegraphics[scale=.45]{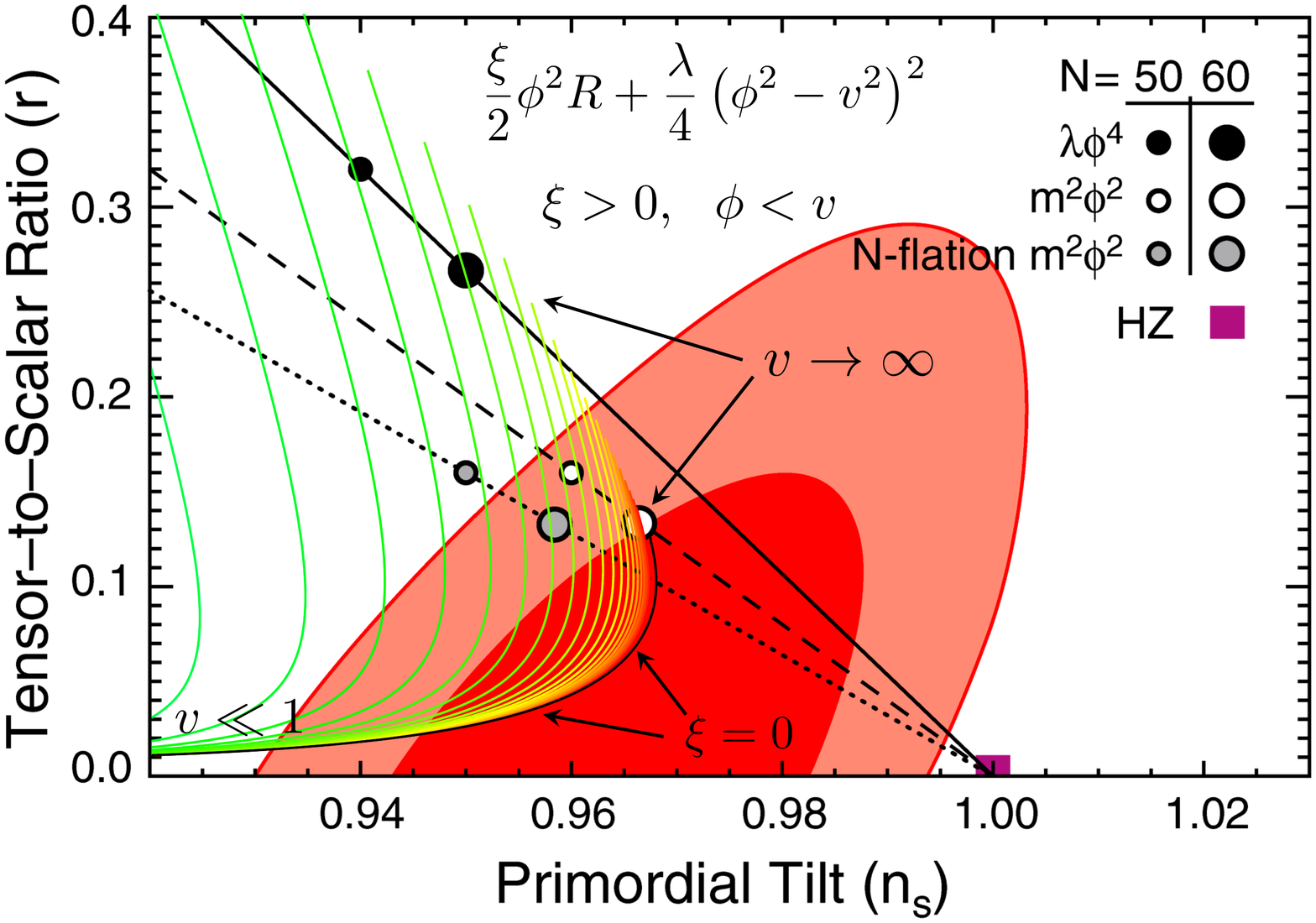}
\caption{$(n_s,r)$ at $N=60$ for the quartic potential \eq{V4}  with positive $\xi$ in the case with inflation at $\phi < v$. $\xi$ increases from right to left and on each curve $v$ increases to $\infty$ as we move away from the bottom-left corner. }
\label{fig:nr4-pos-sf}
\end{figure}

Like the quadratic potential there are different behaviors depending on the sign of $\xi$.  Let us first consider $\xi>0$ where there are two possibilities: `large-$\phi$' and `small-$\phi$' corresponding to $\phi>v$ and $\phi<v$; see \fig{fig:V4}.  In both cases the field rolls to $\phi=v$ but in the small-$\phi$ case from left and in the large-$\phi$ case from right.  Since neither the potential nor the rolling mass $Z^{-1}$ are symmetric around $\phi=v$ we don't expect to obtain the same $(n_s,r)$.  In fact, on the $n_s$-$r$ plane we find that for any fixed $\xi$ there are two branches on each of which $v$ runs from zero to infinity.  This is depicted in \fig{fig:nr4-pos-lf} and \ref{fig:nr4-pos-sf}.

For $\xi=0$ the two branches are joined at $v\to\infty$ by the point corresponding to the $m^2\phi^2/2$ potential, since in this limit the potential near its minimum is indistinguishable from a quadratic one which is symmetric when approached from right or left.  This is not the case for larger $\xi$ because the rolling mass $Z^{-1} \propto \phi^{-2}$ is not symmetric around $\phi=v$ so the field rolls with different velocities on the right and the left.  This can also be seen by looking directly at the potential $V(\varphi)$ written in terms of the field with canonical kinetic term:  If we expand $V(\varphi)$ at large $v$ around its minimum, we find that we can approximate it by a quadratic polynomial only within an interval $\Delta \varphi \ll \sqrt{\frac{1+6\xi}{\xi}}$.  Beyond $\Delta \varphi$ a cubic term dominates which spoils the symmetry of $V(\varphi)$ around its minimum.  Of course, as $\xi$ tends to zero the size of this symmetric interval grows.  In order to have an interval of $\varphi$ that extends $\sim 15$ (in Planck units $M_{p} = 1$) on either side of the minimum and still remains at least 90\% symmetric (i.e., the ratio of the potential at the two ends of the interval is greater than 0.9) we need a $\xi$ as small as ${\cal O}(10^{-6})$.  So only for very small $\xi$ do the ends of the small- and large-$\phi$ branches meet.  This is in marked contrast with the potential $V(\varphi)\propto(\varphi^2-v^2)^2$ for which $\Delta \varphi \propto v$ and hence is always symmetric, for large $v$, over an interval of size $>2\times15$.  In addition, unlike the $\xi=0$ case, the $v\to\infty$ limit of neither branch falls on the curve of \fig{fig:nr2} (corresponding to the quadratic potential) since $Z^{-1}$ near $\phi=0$ and $\phi=v$ are different.

We see that as $\xi$ increases from 0 the small-$\phi$ branch covers the region above and to the left of the $\xi=0$ curve in Fig. \ref{fig:nr4-pos-sf}.  Most of this area is already ruled out, yet the remaining part covers a significant subset of the allowed region.  Meanwhile the large-$\phi$ branches cover a smaller triangular area which shrinks to the point $(1-\frac{2}{N},\frac{12}{N^2})\approx(0.967,0.003)$ in the limit $\xi\to\infty$, independently of the value of $v$.  This result matches the result obtained in Ref.~\cite{Bezrukov:2007ep} for $\xi \gg 1$ and the special case $v\ll 1$. 

Let us now turn to the case of negative $\xi$.  Again we encounter a singularity at $\phi=|\xi|^{-1/2}$ like the quadratic potential.  But this is more complicated since depending on whether $|\xi|^{-1/2}$ is smaller or larger than $v$ the singularity in $V(\phi)$ is to the left or right of $\phi=v$.  In the either case, we ignore the right side of the singularity because it leads to anti-gravity.  Therefore if we plot $V(\varphi)$ we only see the minimum $V=0$ in the $|\xi|^{-1/2}>v$ case and not in the $|\xi|^{-1/2}<v$ case.  These correspond to the blue and red line in \fig{fig:V4n}, respectively.

Let us first consider the red line of \fig{fig:V4n} where $|\xi|v^2>1$. In this case, the only minimum of the potential is at $\varphi =0$ (ignoring the minimum $V=0$ at large $\phi$ in the antigravity regime). This is a minimum with a huge cosmological constant  $V(0)=\frac{\lambda v^4}{4}$, so this regime is unsuitable for inflation describing our part of the universe.

\begin{figure}[ht!]
\centering
\includegraphics[scale=.45]{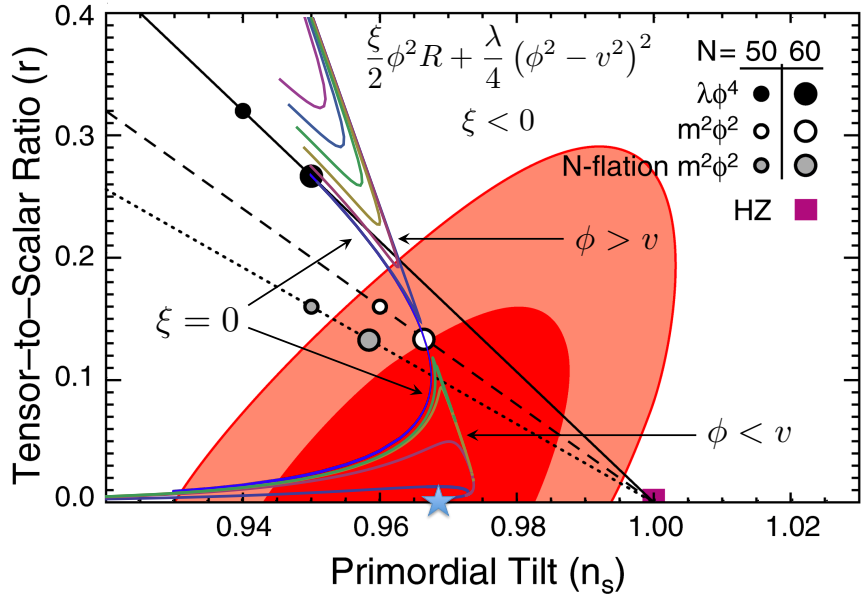}
\caption{
$(n_s,r)$ at $N=60$ for the quartic potential \eq{V4} with negative $\xi$ and $|\xi|v^2<1$ (the blue line of \fig{fig:V4n}).  On each curve $v$ increases from 0 on left to $\lesssim |\xi|^{-1/2}$ on right.  The large-$\phi$ case is shown by the curves above the white circle.  The curves  of constant $\xi$ are shown with $|\xi|$ increasing from bottom to top.  The small-$\phi$ case is represented by the curves below the white circle, with $|\xi|$ increasing from top to bottom.   The cusps correspond to moderately but not too large $v$.  For example, the curve of $\xi=-10^{-6}$ enters the $1\sigma$ region at $v\approx 13$; then continues to the right and for $200\lesssim v \lesssim 800$ stays at the peak of the cusp which is very close to the white circle.  For rather larger $|\xi|$ the curves become almost flat.  For example, the curve of $\xi=-10^{-2}$ crosses the $1\sigma$ region at $v\approx7.8$ with $r\approx0.006$ and continues to the right almost horizontally to $v=9.9$ with $n_s\approx0.972$ and $r\approx0.0069$.  Eventually, in the limit $v\to |\xi|^{-1/2}$, all curves bend to the left and meet at a single point (the star).  Rather unexpectedly, this point coincides with the result of the Higgs inflation model with $\xi \gg 1$ studied in  \cite{Bezrukov:2007ep} (the star in Fig. 6(a)).}
\label{fig:posneg}
\end{figure}

A more interesting possibility $|\xi|v^2<1$ is illustrated by the blue line of \fig{fig:V4n}.  We still have `small-$\phi$' and `large-$\phi$' on the left of the singularity, now defined as $0<\phi<v$ and $v<\phi<|\xi|^{-1/2}$ respectively.  With $N=60$ the result of large- and small-$\phi$ cases are shown in \fig{fig:posneg}.  Unless $|\xi|$ is extremely small, the large-$\phi$ case gives predictions that are incompatible with the WMAP results, see the set of lines above the white circle in  \fig{fig:posneg}.  The small-$\phi$ case is more realistic.  It covers a rather small but observationally important area with $n_s\lessapprox0.97$ and the tensor-to-scalar ratio $r$ as tiny as $\sim10^{-3}$. The motion along the  lines from left to right in \fig{fig:posneg} shows what happens in this model, for a given value of $\xi<0$, when $v$ grows. These curves have some interesting features.

First of all, there are cusps, i.e., maxima just before the right end of each $\xi=\rm const$ curve. This  feature can be qualitatively understood as follows: When $\phi< v \ll |\xi|^{-1/2}$, it doesn't matter whether $\xi$ is positive or negative. In other words, $\phi=v$ (where the potential vanishes) is far away from $\phi=|\xi|^{-1/2}$ where effects of $\xi$ begin to show up.  So the left ends of the curves below the line $\xi =0$ of \fig{fig:posneg} are almost the same as those of \fig{fig:nr4-pos-sf}.  But when $v$ grows and starts approaching $|\xi|^{-1/2}$, then the potential differs for positive and negative $\xi$, since for the latter case a singularity develops.  This is the place when the curves of constant $\xi$ in \fig{fig:nr4-pos-sf} and \fig{fig:posneg} depart; the former go up (because they approach the large-field-excursion-type potential of $m^2 \phi^2/2$) and the latter go down (because they approach a small-field-excursion-type potential similar to new inflation).

The second feature is that, regardless of the value of $\xi$, all of these curves converge on their right end to the same point $(n_s, r) = (1-\frac{2}{N}, \frac{12}{N^2}) = (0.967,0.003)$. This is the limit when the minimum at $\phi_{0}=v$ is very close to the singularity at $\phi=|\xi|^{-1/2}$, i.e., when $|\xi| v^2$ approaches 1 from below.  We now show that it corresponds to a new-inflation type potential with an extremely flat plateau.

\begin{figure}[ht!]
\centering
\subfigure[]{
\includegraphics[scale=.67]{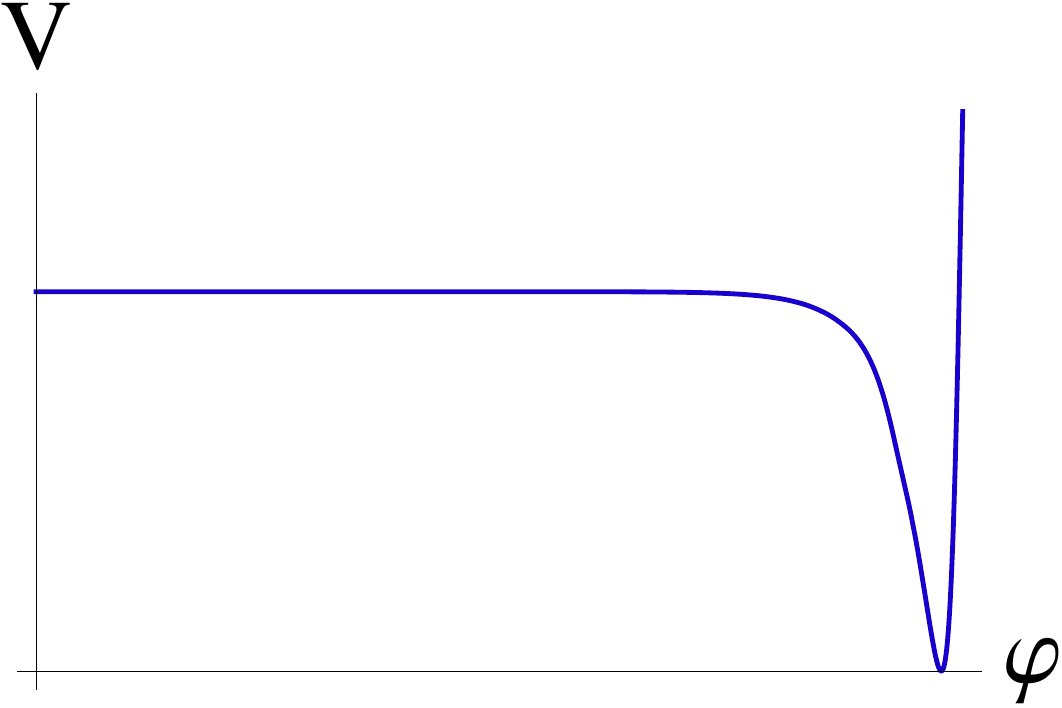}
\label{fig:newlonga}}~~~~~
\subfigure[]{
\includegraphics[scale=.62]{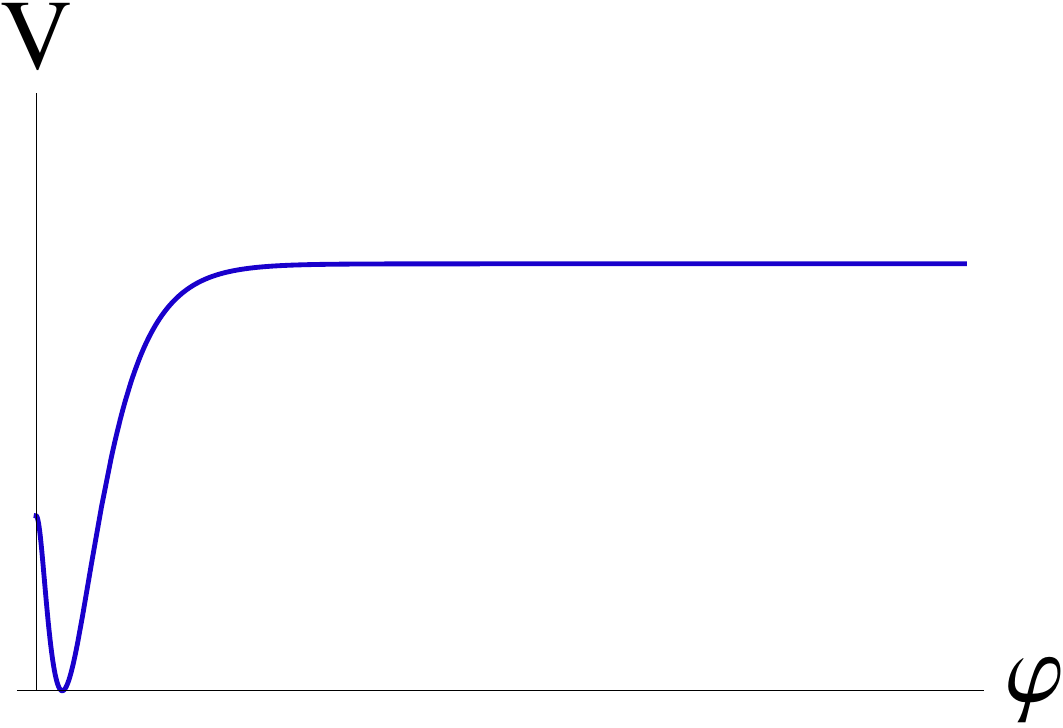}
\label{fig:newlongb}}
\caption{The potential ${\lambda\over 4} (\phi^2-v^2)^2$ as a function of the canonically normalized field $\varphi$ in the Einstein frame.  (a) $\xi < 0$, $1-|\xi|v^{2} \ll 1$. (b) $\xi > 0$.  Notice the similarity between these two potentials. However, physical interpretation of these two potentials is very different. The value of the canonically normalized field $\varphi$ in the model with $\xi < 0$ typically is very large. Meanwhile the value of this field for the model with $\xi > 0$ can be very small, which is the basis of the Higgs inflation scenario  \cite{Bezrukov:2007ep}.}
\end{figure}

In this limit we can use \eq{negative} to approximate the form of the potential in the vicinity of its minimum at $\phi_{0}=v$:
\begin{equation}\label{asympot}
V(\varphi) = \left(\frac{\lambda}{4\xi^{2}}\right) \left( 1 - \exp \Bigl(-\sqrt{\frac{2}{3}} (\varphi_{0} - \varphi) \Bigr) \right)^2,
\end{equation} 
where $\varphi_0$ is the location of the minimum in terms of the canonical field $\varphi$. One can show that during inflation, in the slow-roll regime, $ \exp \Bigl(-\sqrt{\frac{2}{3}} (\varphi_{0} - \varphi) \Bigr) \ll 1$.  Therefore by replacing $\varphi_{0} - \varphi \to \phi$ one can represent this potential during inflation as 
\begin{equation}\label{BSpot}
V =\frac{\lambda}{4\xi^{2}} \left( 1 + \exp \Bigl(-\sqrt{\frac{2}{3}} \phi \Bigr) \right)^{-2}.
\end{equation}
At large $\varphi$, this potential coincides with the inflationary potential  in the Higgs inflation scenario with the potential ${\lambda\over 4} (\phi^2-v^2)^2$ with $v \ll 1$ and $\xi \gg 1$ in Ref.~\cite{Bezrukov:2007ep}. To illustrate this statement, one may compare the potential ${\lambda\over 4} (\phi^2-v^2)^2$ with $\xi <0$ in the limit $|\xi|v^{2}\to 1$ with the potential in the Higgs inflation model \cite{Bezrukov:2007ep} with $\xi > 0$ and $v < 1$, for the same values of parameters $\lambda$ and $|\xi|$, see  Figs. \ref{fig:newlonga} and   \ref{fig:newlongb}. The results of our calculations of the parameters $n_{s}$ and $r$ and of the amplitude of perturbations of metric in this model coincide with the corresponding results of Ref.~\cite{Bezrukov:2007ep}.

The relation between the inflationary regime with $\phi < v$ and $\xi < 0$ in the limit $|\xi|v^2\to 1$ and Higgs inflation with $\phi > v$ and  $\xi \gg 1$  is quite interesting and even somewhat mysterious, as if there is some hidden duality between these two classes of models with opposite signs of $\xi$. Moreover, the same set of the cosmological parameters $(n_s, r) = (1-\frac{2}{N}, \frac{12}{N^2}) = (0.967,0.003)$ also appears in Starobinsky model \cite{Starobinsky:1979ty}, see \cite{Mukhanov:1981xt,Starobinsky:1983zz}.

The value of $\lambda$ can be determined from the observed value of amplitude of density perturbations $\delta_\rho \approx 5\times10^{-5}$.  For the two observationally interesting cases the situation is as follows:  For the large-$\phi$ case with positive $\xi$, $\lambda$ decreases along the constant-$\xi$ lines.  At $\xi=0$ we have $\lambda\approx2\times10^{-13}$ and for large $v$ we have $\lambda\approx4\times10^{-11}/v^2$ (this follows from the fact that for large $v$ the quartic potential near its minimum looks like a quadratic one with mass squared $m^2=\lambda v^2$).  For large $\xi$ we have $\lambda\approx5\times10^{-10}\xi^2$ independently of $v$, as in Ref.~\cite{Bezrukov:2007ep}.  For the small-$\phi$ case with negative $\xi$ the behavior is similar.  For small $|\xi|$ and large $v$ (but not too large to pass the cusp peak) we have $\lambda\approx4\times10^{-11}/v^2$.  In the limit $|\xi| v^2 \to 1$ the potential is the same as that of Ref.~\cite{Bezrukov:2007ep} and hence again we have $\lambda\approx5\times10^{-10}\xi^2$.  Thus we see that $\lambda$ can be parametrically large or small.

\section{Conclusions}

In this paper we studied the simplest chaotic inflation models with potentials ${m^{2}\over 2}\phi^{2}$ and ${\lambda\over 4}(\phi^{2}-v^{2})^{2}$, with the field $\phi$ nonminimally coupled to gravity, for arbitrary values of the parameters $v$ and $\xi$. We demonstrated that these models can be implemented in the context of supergravity, along the lines of \cite{Kallosh:2010ug}, and the inflationary trajectory can be stabilized by a proper choice of the \K\, potential. 

We found that by changing parameters of these models one can cover a significant part of the space of the observable parameters ($n_{s},r$) allowed by recent observational data. As one can see from Figs. 4, \ref{fig:nr4-pos-lf} and \ref{fig:posneg}, the best fit is achieved by the model ${m^{2}\over 2}\phi^{2}$ with $\xi > 0$, and by the model ${\lambda\over 4}(\phi^{2}-v^{2})^{2}$ in two different regimes: $\xi > 0$, with inflation at $\phi > v$, and $\xi < 0$, with inflation at $\phi < v$. 

In particular, Fig. 4 shows that by adding a term ${\xi\over 2}\phi^{2}R$ with $\xi \sim 10^{{-2}}$ one can suppress the tensor/scalar ratio $r$ for the model ${m^{2}\over 2}\phi^{2}$ from $r\sim 0.13 - 0.15$ (depending on $N$) down to $r \sim 0.01$. Similarly, investigation of the model ${\lambda\over 4}(\phi^{2}-v^{2})^{2}$, which is ruled out by observations for $v< 1$, shows that by adding to this theory a term ${\xi\over 2}\phi^{2}R$ with $\xi \gtrapprox 0.0027$  one makes it consistent with observational data within 1$\sigma$ (red area in Fig. \ref{fig:nr4-pos-lf}), for all values of $v$. In other words, in order to make this model consistent with observations, there is no need to go to the limit $\xi \gg1$; a very small positive value of $\xi$ is quite sufficient. Further increase of $\xi$ leads to a rapid convergence of the predictions to a single point $n_{s}=0.967$ and $r =0.003$, independently of the value of $v$, see the gray star in Fig. \ref{fig:nr4-pos-lf}. 

One of  the most unexpected results of our paper is the coincidence between the inflationary predictions of the model ${\lambda\over 4}(\phi^{2}-v^{2})^{2}$ for $\xi \gg 1$, as shown in Fig. \ref{fig:nr4-pos-lf}, and of the same model with $\xi <0$ in the limit $|\xi|v^{2}\to 1$, Fig. \ref{fig:posneg}. In both cases one finds $n_{s}=0.967$ and $r =0.003$.
 
 Thus, the non-minimal coupling of the inflaton field to gravity may significantly modify predictions of the simplest inflationary models. In several cases, these modifications suppress the amplitude of tensor modes to the level below $r \sim 10^{{-2}}$, which may make tensor modes difficult to observe. In addition, a proper theoretical interpretation of the cosmological data may become difficult because one can obtain the same set of parameters ($n_{s},r$) in different models. This is a general problem discussed e.g. in \cite{Easson:2010uw}. The degeneracy between the predictions of different models in certain cases may be removed if one takes into account additional scalar fields which may exist in the supergravity-based versions of these models. For a certain choice of the \K\, potential, some of these fields may become light  along the inflationary trajectory. This may lead to an additional source of inflationary perturbations for some of these models, which can be very non-gaussian \cite{Demozzi:2010aj}. Thus, this class of models has some unique features, which provide new interesting possibilities for a proper interpretation of upcoming observational data.

\begin{acknowledgments}
The authors are grateful to R.~Kallosh, T.~Rube and V.~Vanchurin for useful comments.  The work by A.L. and M.N. was supported by NSF grant PHY-0756174.  M.N.'s work was supported in part by a Mellam Family Fellowship.  The research of A.W. was supported in part by NSF grant PHY-0756174, as well as by the Impuls und Vernetzungsfond of the Helmholtz Association of German Research Centres under grant HZ-NG-603.
\end{acknowledgments}

\end{document}